\documentstyle[prl,aps,multicol,epsf]{revtex}

\DeclareMathAlphabet{\mathitbf}{T1}{cmr}{bx}{it}
\newcommand{\vecx}{{\mathitbf x}}
\newcommand{\vecy}{{\mathitbf y}}
\newcommand{\vecs}{{\mathitbf S}}
\newcommand{\vecr}{{\mathitbf R}}
\begin{document}

\draft

\title{Spin Glass Ordering in Diluted Magnetic Semiconductors: a Monte Carlo
Study}

\author{E. Marinari$^1$, V. Mart\'{\i}n-Mayor$^1$, A. Pagnani$^2$}

\address{$^1$ Dipartimento di Fisica,
Universit\`a di Roma ``La Sapienza'',
P. Aldo Moro 2, 00185 Roma (Italy)\\
INFN Sezione di Roma - INFM Unit\`a di Roma\\
$^2$ Dipartimento di Fisica and INFM,
Universit\`a di Roma ``Tor Vergata'',
V. della Ricerca Scientifica 1, 00133 Roma (Italy)
}

\date{\today}
\maketitle

\begin{abstract}
We study the temperature-dilution phase diagram of a site-diluted
Heisenberg anti-ferromagnet on a fcc lattice, with and without the
Dzyaloshinskii-Moriya anisotropic term, fixed to realistic microscopic
parameters for $IIB_{1-x}\,Mn_x\,Te$ ($IIB$$=$$Cd$, $Hg$, $Zn$).  We
show that the dipolar Dzyaloshinskii-Moriya anisotropy induces a
finite-temperature phase transition to a spin glass phase, at
dilutions larger than 80$\%$. The resulting probability distribution
of the order parameter $P(q)$ is similar to the one found in the cubic
lattice Edwards-Anderson Ising model. The critical exponents undergo
large finite size corrections, but tend to values similar to the ones
of the Edwards-Anderson-Ising model.
\end{abstract}

\begin{multicols}{2}
\narrowtext

Although most theoretical investigations on spin glass (SG) systems
\cite{BOOKS} have focused on models based on Ising spins, systems that
have been investigated in the recent period, like the diluted magnetic
semiconductors $IIB\,Mn\,Te$ series, $IIB=Cd,Hg,Zn$, i.e.
$Cd_{1-x}\,Mn_x\,Te$ \cite{EXPMNTE_Cd,EXPMNTE_Hg},
$Hg_{1-x}\,Mn_x\,Te$ \cite{EXPMNTE_Hg} and $Zn_{1-x}\,Mn_x\,Te$
\cite{EXPMNTE_Zn} (and somehow even typical experimental samples
\cite{EXP} like $Cu\,Mn$, $Ag\,Mn$, $Eu_x\,Sr_{1-x}\,S$), are closer
in nature to continuous Heisenberg spins.  Early computer simulations
suggested that in three spatial dimensions neither systems with local
interactions and Heisenberg \cite{MORRIS,MACMILLAN,OLIVE} or XY
\cite{MORRIS,JAIN} spins, nor systems with long-range RKKY
interactions \cite{RKKY} undergo a finite temperature SG phase
transition.  This fact could be potentially annoying from a
phenomenological point of view, but it becomes acceptable after
noticing that a small anisotropic interaction, neglected in the above
calculations, could induce a finite temperature SG phase transition.

Still the situation is not crystal clear: the work of \cite{ANGULO}
claimed that the most important anisotropic coupling in $IIB\,Mn\,Te$
materials \cite{LARSON}, the Dyalozhinskii-Moriya (DM) interaction, is
not able to induce a SG phase.  On the contrary a theoretical analysis
\cite{MAUGER} of experimental data on the $IIB\,Mn\,Te$ series was used
to suggest the presence of SG ordering in $3D$ Heisenberg spin glasses
(for finite temperature and no anisotropies): recent numerical
simulations \cite{KAWAMURA} support the existence of a chiral phase
transition in such systems.  The role of the anisotropy was
reconsidered in \cite{MATSUBARA}, where the Heisenberg spin
Edwards-Anderson (HEA) model was considered with the addition of a
random pseudo-dipolar interaction: clear signatures of a finite
temperature SG phase were found.  This result is not consistent with
the one of \cite{ANGULO} (that was considered as being based on a
realistic modelization of $IIB\,Mn\,Te$, even if the direction of the
DM vectors, see (\ref{HAMILTONIAN}), was chosen at random, while they
should be periodic along the lattice \cite{LARSON}).  We remind at
last that a diluted Ising anti-ferromagnet on a fcc lattice has been
studied in \cite{WENGEL}, where the signature of a SG phase transition
for low enough densities has been detected.

We have taken here the point of view of trying to be as realistic as
possible, analyzing a model as close as possible to the experimental
samples. We show that non-random DM terms (selecting a realistic value
for the anisotropy) are able to induce a SG
phase transition in a $3D$ Heisenberg spin glass on a fcc lattice.  We
analyze and discuss in detail the values of critical exponents: the
experimental results for the $IIB_{1-x}\,Mn_x\,Te$ materials
\cite{EXPMNTE_Cd,EXPMNTE_Zn} are in good agreement with the most
accurate calculations for the Edwards-Anderson model with Ising spins
(IEA) on the cubic lattice~\cite{PALASSINI} ($\nu=1.8 \pm 0.2)$,
$\eta=-0.26 \pm 0.04$), but the numerical simulations of \cite{WENGEL}
and of \cite{MATSUBARA}, yielded $\nu\approx 1.0$.  We will show that
the numerical calculation of the critical exponents on the accessible
lattice sizes suffer from serious finite-size corrections, and that a
systematic analysis of the numerical data establishes clear trend
towards values of $\nu$ larger than the ones found in
\cite{MATSUBARA,WENGEL}, and close to the experimental values.

The site-diluted anti-ferromagnetic (AFM) Heisenberg model on the fcc
lattice, with and without DM anisotropy, is a model for the
$IIB_{1-x}\,Mn_x\,Te$ series, where the $Mn$ atoms form an fcc lattice
with localized (Heisenberg) spins interacting through short-range
(super-exchange) AFM terms, while the magnetically inert $IIB$ atoms
randomly replaces the $Mn$ over the lattice.  An AFM interaction on
the fcc lattice is frustrated, and gives rise to some interesting
order-disorder phenomena \cite{VILLAIN}, both with Heisenberg
\cite{O3A} and Ising spins \cite{ISINGAFfcc}.  The dilution disorder
deletes some of the sites on the system, thus providing the random
combination of frustrated and unfrustrated plaquettes, that is
believed to be essential for SG ordering.  The Hamiltonian of the
system is

\begin{equation}
H=J \sum_{\langle \vecx,\vecy\rangle'}\,
\left[ \vecs_{\vecx}\cdot\vecs_{\vecy}+
\frac{D}{J} \vecr_{\vecx-\vecy}\cdot(\vecs_{\vecx}\wedge\vecs_{\vecy})\right]
\,,
\label{HAMILTONIAN}
\end{equation}
where the fields $\vecs=(S_1,S_2,S_3),\,$ $S_1$, $S_2$ and $S_3$ real
with $\vecs^2=1$, represent the spin of the $Mn$ atoms, and $J>0$.  A
lattice site is randomly occupied by a spin with probability $p$. The
sum labeled by $\langle \vecx,\vecy\rangle'$ runs over the pairs of
occupied nearest neighboring sites of the lattice.  The unit-length
vectors $\vecr_{\vecx-\vecy}$ specify the DM anisotropy, and they
verify $\vecr_{\vecx-\vecy}=-\vecr_{\vecy-\vecx}$. Following
\cite{LARSON} we set \hbox{$ \vecr_{(\frac{1}{\sqrt{2}},
\frac{1}{\sqrt{2}},0)}= (\frac{1}{\sqrt{2}}, -\frac{1}{\sqrt{2}},0)$},
while the other five independent vectors are obtained using the
three-fold rotation symmetries of the lattice. In Ref.~\cite{LARSON}
the ratio $D/J$ has been estimated to be $0.054$ for
$Zn_{.77}\,Mn_{.33}\,Te$ and $Cd_{.77}\,Mn_{.33}\,Te$, and is very
mildly dependent on the composition of the sample: we have fixed it to
$0.06$ for simplicity. As the local magnetic field acting on spin
$\vecs_\vecx$ is (see (\ref{HAMILTONIAN}))

\begin{equation}
h_\vecx =J \sum_{\vecy,\,||\vecx-\vecy||=\sqrt{2}}\, \vecs_\vecy+
\frac{D}{J} (\vecs_{\vecy}\wedge\vecr_{\vecx,\vecy})\,,
\end{equation}
it is easy to implement a heat-bath algorithm and the over-relaxed
micro-canonical algorithm of \cite{O3A}.  We have found that the
combination of these two updates tremendously reduces the
thermalization effort. In the production runs we have performed a
full-lattice heat-bath sweep followed by $19$ over-relaxed updates,
that will be referred in the following as an elementary Monte-Carlo
step (EMCS).  In order to define the observables, it is useful to
consider a {\em replica} (i.e. a thermally independent system
with the same set of occupied sites, that we denote
$\tilde{\vecs}(\vecx)$). The measured observables can be most easily
described in terms of the following three basic fields: the tensor
field ($\tau_{\alpha,\beta}(\vecx)= S^\alpha_\vecx
S^\beta_\vecx-\frac{1}{3}\delta^{\alpha,\beta}$, if the lattice site
$\vecx$ is occupied, and zero otherwise), the tensorial overlap
($O_{\alpha,\beta}(\vecx)=S^\alpha_\vecx {\tilde S}^\beta_\vecx$), and
the scalar overlap field ($q(\vecx)=
\vecs_{\vecx}\cdot\tilde{\vecs}_{\vecx}$).

The rationale for studying the tensor field $\tau(\vecx)$
\cite{RP2AFM} is that previous studies of AFM diluted systems on the
fcc lattice \cite{WENGEL}, showed that only for moderate dilutions the
system ceases to develop AFM ordering. The tensorial magnetization is
an ideal order parameter to check this possibility, since it will be
non-vanishing for {\em any} conceivable type of AFM or helicoidal
ordering. Moreover, it would also work on more sophisticated, yet
trivial situations like the found in $D=0$, $p=1$ \cite{O3A}. The
tensorial overlap \cite{BARBARA} is most adequate to study the
isotropic ($D=0$) case, since in this situation the Hamiltonian
posses a $O(3)$ global symmetry: when the anisotropy is
switched on the symmetry reduces to $Z_2$, and the use of the scalar
overlap becomes natural. For all three fields, one can
define straightforwardly the corresponding susceptibility, Binder
parameter and a finite lattice correlation length
\cite{KAWAMURA,PALASSINI,RP2AFM}.

\begin{figure}
\epsfxsize=0.95\columnwidth
\epsffile{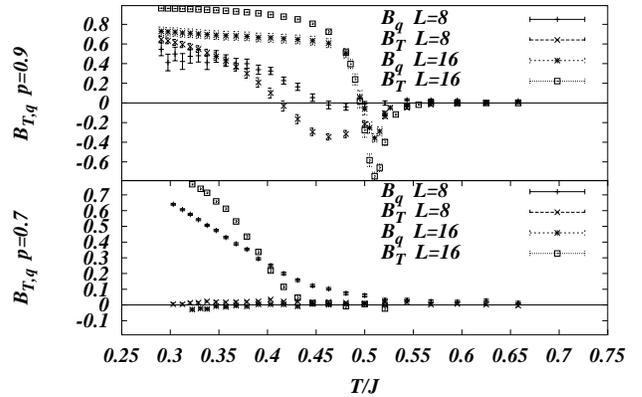}
\caption{Tensorial ($B_T$) and scalar ($B_q$)
overlap Binder parameter versus temperature
for sizes $L=8,16$ and dilution $p=0.9$ (upper part) and $p=0.7$
(lower part).}
\label{fig1}
\end{figure}

The model (\ref{HAMILTONIAN}) without impurities ($p=1.0, D=0.06J$)
undergoes a phase transition at $T_{\mathrm{c}}(p=1)\approx 0.60J$
from a paramagnetic phase to an AFM phase, as shown by the behavior of
the tensorial magnetization. For larger dilutions a lower temperature
value needs to be reached in order to exit from the paramagnetic
phase: the critical line, $T_{\mathrm{c}}(p)$, will eventually reach
zero temperature at the percolation threshold for the magnetic ions
($p_{\mathrm{c}}\approx 0.2$).  The first question is for which
dilution the system forgets its global AFM ordering. In order to
answer this question we have performed slow annealings in $60$ samples
(and its corresponding replicas) at dilutions $p=1.0,0.9$ and $0.8$,
in lattices $L=8$ and $16$; at $p=0.7$ and $p=0.6$ (that will be shown
to be in the SG compositional range), we have annealed $700$ samples.
The results for the Binder cumulant of the tensor and scalar overlap
fields at $p=0.7$ and $p=0.9$ are displayed in figure~\ref{fig1}. At
$p=0.9$ for both observables we find a low temperature, AFM ordered
phase, since the tensorial magnetization is non-vanishing. There is a
strong dip close to the phase transition point, which probably is very
plausibly of first order. On the contrary for $p=0.7$ it is clear that
the tensorial magnetization is no longer an appropriate order
parameter. For $p=0.8$ (not shown in the plot), our results indicate a
cross-over regime between the two situations. Therefore, the low
temperature phase turns from AFM to SG at $0.7<p_{\mathrm{c}}<0.8$,
similarly to what happens in the Ising case \cite{WENGEL}. Also for
$D=0,p<0.8$ we do not find an AFM ordered phase.

\begin{figure}
\epsfxsize=0.9\columnwidth
\epsffile{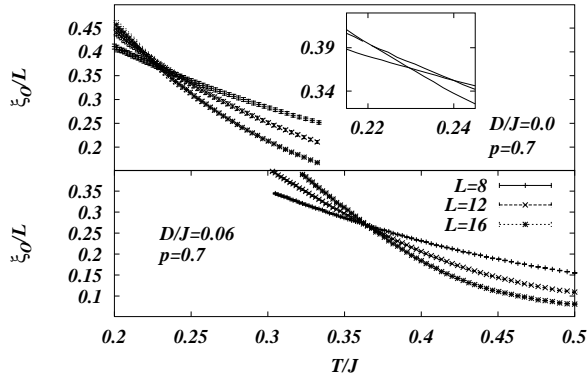}
\caption{Correlation length of tensorial overlap (in units of lattice
size) vs. temperature for $D=0$ (upper
part and zoom) and in presence of anisotropy $D=0.06J$ (lower part).}
\label{fig2}
\end{figure}

In order to quantify how strong the effect of the anisotropy is, we
compare the system at $p=0.7$ with $D=0$ and $D=0.06J$. In
figure~\ref{fig2} we show the correlation length of the tensorial
overlap in units of the lattice size. This operator should be zero in
the paramagnetic phase, diverging in the SG phase, and at the critical
point reaches a finite universal value.  In the isotropic case (see
zoom in upper part of figure~\ref{fig2}) the crossing point of the
$L=12$ and $16$ lattices is not clearly resolved, and their respective
crossings with the $L=8$ curve shifts to lower temperature with
growing lattice size. Moreover, the data (not shown in the plot) for
the Binder Cumulant of the tensorial overlap rapidly grow at the
crossing temperature of the correlation length, but then saturate at a
value which decreases with the lattice size, without a crossing,
similarly to the results shown in \cite{KAWAMURA}, where it has been
shown that the chiral-glass phase appears precisely at the
temperatures at which the Binder cumulant of the tensorial overlap
grows.

On the other hand, with a $6\%$ DM anisotropy (see the lower frame of
figure~\ref{fig2}), we find a neat crossing of the correlation length
(the tensorial overlap Binder cumulant has a marked dip, in contrast
with the scalar overlap shown in the lower frame of
figure~\ref{fig1}). As the phase transition for the anisotropic system
occurs at a temperature $80\%$ higher than the one close to the
crossings of the $D=0$ case, that according to \cite{KAWAMURA} signal
a real chiral-glass phase transition, the natural conclusion is that
the DM anisotropy is {\em not} a smooth perturbation that reveals a
hidden chiral-glass ordering.

\begin{figure}
\epsfxsize=0.95\columnwidth
\epsffile{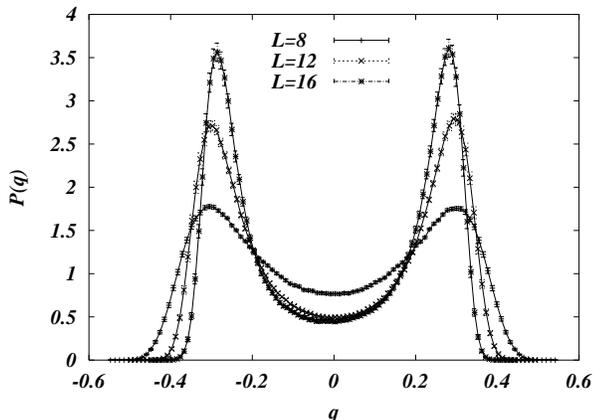}
\caption{Probability distribution function of scalar overlap
for $L=8,12,16$ at $p=0.6$, $T=J/4.5 \approx 0.78
T_{\mathrm{c}}$ and $D=0.06J$.}
\label{fig3}
\end{figure}

In order to characterize more precisely the SG phase we have studied
the distribution of the scalar overlap at $p=0.6, D=0.06 J$ and
$T=J/4.5\approx 0.78 T_{\mathrm{c}}$. At this temperature we have
estimated the mean thermalization time in the $L=16$ lattice, by
considering a logarithmic plot of the mean overlap susceptibility of
$64$ samples, as a function of MC-time, starting from a random
configuration, and we have found it to be of order $250$ EMCS. After
that we have performed a run with $800$ samples, with $L=8,12,16$,
performing $8000$ EMCS on each sample, and taking a measure every $4$
EMCS.  We display $P(q)$ in figure~\ref{fig3}: is central part is
remarkably stable for growing lattice size. Therefore, on the lattice
sizes that we are able to thermalize, the pattern we obtain is
completely analogous to the one found for the Ising EA model in $3D$
\cite{JSP}.

Due to the global $Z_2$ symmetry of the Hamiltonian
(\ref{HAMILTONIAN}), one would expect it to belong to the same
universality class of the IEA model in $3D$, which seems even more
plausible from our measures of the $P(q)$. To further investigate this
relation we have measured the critical exponents, in the dilution
range where we definitively find SG ordering, namely $p=0.7$ and
$0.6$.  Since we have at our disposal only a narrow range of lattice
sizes, it is important to use a finite size scaling analysis that
allows to study the scaling corrections.  We have used the quotient
method of \cite{RP2AFM}, that has been particularly useful in the
study of scaling corrections in disordered systems \cite{ISDIL}.  
We
measure an operator $O$, diverging in the infinite size limit at
criticality as $|T-T_{\mathrm{c}}|^{-x_O}$, on two finite lattices of
sides $L$ and $sL$, 
and we select the temperature value where
the two correlation
lengths in units of the lattice size coincide
(see the crossing of figure \ref{fig2}).
For the quotient of
these two measures we have
\begin{equation}
\left.\frac{O(sL,T)}
{O(L,T)}\right|_{\frac{\xi(sL,T)}{\xi(L,T)}=s}=
s^{x_O/\nu}\left(1+{\cal O}(L^{-\omega})\right)\,,
\label{QUOTIENT}
\end{equation}
where $\omega>0$ is related to the first irrelevant operator in the
Renormalization Group sense. The main advantage of the relation
(\ref{QUOTIENT}) is that the large statistical correlation between the
measurements of $O$ and $\xi$ allows to measure the quotient with
sufficient accuracy as to uncover the scaling corrections. In our
$Z_2$ symmetric case we have of course used the scalar overlap
correlation length.  Our results are displayed in table
\ref{table:exponentes}, and they do show the presence of significant
scaling corrections ($\nu$ is computed using the temperature
derivative of $\xi_q$, $\eta$ from the susceptibility of the scalar
overlap). Since we only have few lattice sizes it is meaningless to
try an infinite size extrapolation as the one of \cite{ISDIL}.  We
should still mention that the critical exponents obtained by a simple
log-log fit (for instance, with data measured at the maximum of the
specific-heat) is roughly equivalent to an average of the transient
exponents displayed in our table. Therefore the value of $\nu\approx
1$ found with Ising spins \cite{WENGEL} or in the HEA model with
pseudo-dipolar anisotropy \cite{MATSUBARA}, is most probably a
preasymptotic value. In fact, the best available results for the IEA
model in $3D$\cite{PALASSINI}, $\nu=1.8(2), \eta=-0.26(4)$ are
plausible infinite volume extrapolations for our results. However, in
order to definitively elucidate this point, it would be helpful to use
the extrapolation method of \cite{SOKAL,PALASSINI}, that allows to
work in the paramagnetic region with a significantly smaller
thermalization effort.

We have shown for the first time that dipolar DM anisotropic local
interactions are able to induce a SG phase transition for Heisenberg
spins in three dimensions. This result has been obtained with the very
small realistic value of the anisotropy coupling constant in the
$IIB\,Mn\,Te$ series. Given the dramatic effect of this small
perturbation term in the Hamiltonian, we suggest that the chiral-glass
mechanism proposed in \cite{MAUGER} is overwhelmed by the neglected DM
term.  We have studied the temperature-dilution phase diagram of the
Hamiltonian (\ref{HAMILTONIAN}) in a large dilution range. We have
found that the low-temperature phase changes from AFM to SG order
between $p=0.8$ and $p=0.7$.  We have used a combination of
micro-canonical and heat-bath Monte Carlo update, that have allowed us
to thermalize a $L=16$ lattice at $T=0.78 T_{\mathrm{c}}$, in the SG
phase. We have measured the distribution of the overlap, finding
results analogous to the ones of Ising EA $3D$ model on similar
lattice sizes.  We have given an estimate the critical-exponents on
the SG dilution range. Our results suffer from severe finite size
corrections, but it is plausible to deduce that critical exponents are
converging to the Ising EA results, as the experimental results for
the $IIB\,Mn\,Te$ suggest \cite{EXPMNTE_Cd,EXPMNTE_Zn}. Further open
questions need clarification: a better measure of the critical
exponents using the method of \cite{SOKAL}, the precise
characterization of the AFM phase, and a detailed study of the order
of the paramagnetic anti-ferromagnetic phase transition at low
dilution.

We acknowledge interesting discussions with
D. Brogioli, B. Coluzzi, A. Geddo-Lehmann,
G. Parisi, F. Ricci-Tersenghi, J. J. Ruiz-Lorenzo.
V.M.M. is a M.E.C. fellow and has been partially supported by
CICyT(AEN97-1708 and AEN99-1693).  The simulations have been performed
using the Pentium clusters of the Universit\`a di Cagliari (Kalix2)
and the Universidad de Zaragoza (RTNN collaboration).

\vbox{ \narrowtext
\begin{table}[b]
\begin{tabular}{rllll}
$(L_1,L_2)$ & {$\nu (p=0.7)$} & {$\nu(p=0.6)$} & {$\eta(p=0.7)$} & 
{$\eta(p=0.6)$}\\\hline
(8,12) & 0.941(9)  &  1.08(2) &  0.443(6)& 0.193(8)\\
(8,16) & 1.055(13) &  1.32(2) &  0.392(3)& 0.0959(12)\\
(12,16)& 1.277(22) &  1.91(7) &  0.28(6)& \hskip-3pt -0.10(5)\\
\end{tabular}
\caption{Transient critical exponents $\nu{(L_1,L_2)}$ and $\eta{(L_1,L_2)}$
obtained with the quotient method (see text).}
\protect\label{table:exponentes}
\end{table}
}

\end{multicols}
\end{document}